\documentclass[prb,twocolumn,balancelastpage,superscriptaddress,floatfix,showpacs,letterpaper]{revtex4-1}
\usepackage{times}
\usepackage{verbatim}
\usepackage{bm}
\usepackage{bbm}
\usepackage{makeidx}
\usepackage{pifont}
\usepackage{amsmath}
\usepackage{amsfonts}
\usepackage{amscd}
\usepackage{amsthm}
\usepackage{amssymb}
\usepackage{braket}
\usepackage{graphicx}
\graphicspath{{/}}
\usepackage{longtable}
\usepackage{dcolumn}

\newcommand{\diag}                      {\mathop{\mathrm{diag}}\nolimits}
\newcommand{\ee}                        {e}                                  
\newcommand{\idmat}                     {\mathbbm{1}}                        
\newcommand{\ii}                        {i}                                  
\renewcommand{\vec}[1]                  {\mathbf{#1}}                        
\newcommand{\unitvec}[1]                {\vec{e}_{#1}}                       
\newcommand{\avg}[1]                    {\langle#1\rangle}                   

\providecommand{\eqn}                   {Eq.~}
\providecommand{\eqns}                  {Eqs.~}

\newcommand{\coloronline}               {(Color online)}

\newcommand    {\vk}         {\vec{k}}
\newcommand    {\lB}         {l_\mathcal{B}} 
\newcommand    {\muB}        {\mu_\mathrm{B}} 
\newcommand    {\kB}         {k_\mathrm{B}} 
\newcommand    {\magnB}      {\mathcal{B}} 
\newcommand    {\sigmaH}     {\sigma_\mathrm{H}} 
\newcommand    {\sigmaspH}   {\sigma_\mathrm{H}^{\mathrm{sp}}} 
\newcommand    {\tildegE}    {\tilde{g}_\mathrm{E}} 
\newcommand    {\tildegH}    {\tilde{g}_\mathrm{H}} 
\newcommand    {\chiE}       {\chi_\mathrm{E}} 
\newcommand    {\chiH}       {\chi_\mathrm{H}} 
\newcommand    {\lambdaex}   {\lambda_\mathrm{ex}} 

\begin{document}
\title{Reentrant topological phases in Mn-doped HgTe quantum wells}
\author{W. Beugeling}
\affiliation{Institute for Theoretical Physics, Utrecht University, Leuvenlaan 4, 3584 CE Utrecht, The Netherlands}
\author{C. X. Liu}
\affiliation{Physikalisches Institut (EP3), Universit\"at W\"urzburg, Am Hubland, 97074 W\"urzburg, Germany}
\affiliation{Department of Physics, Pennsylvania State University, University Park, Pennsylvania 16802, USA}
\author{E. G. Novik}
\affiliation{Physikalisches Institut (EP3), Universit\"at W\"urzburg, Am Hubland, 97074 W\"urzburg, Germany}
\author{L. W. Molenkamp}
\affiliation{Physikalisches Institut (EP3), Universit\"at W\"urzburg, Am Hubland, 97074 W\"urzburg, Germany}
\author{C. \surname{Morais Smith}}
\affiliation{Institute for Theoretical Physics, Utrecht University, Leuvenlaan 4, 3584 CE Utrecht, The Netherlands}

\date{\today}

\begin{abstract}
Quantum wells of HgTe doped with Mn display the quantum anomalous Hall effect due to the magnetic moments of the Mn ions. In the presence of a magnetic field, these magnetic moments induce an effective nonlinear Zeeman effect, causing a nonmonotonic bending of the Landau levels. As a consequence, the quantized (spin) Hall conductivity exhibits a reentrant behavior as one increases the magnetic field. Here, we will discuss the appearance of different types of reentrant behavior as a function of Mn concentration, well thickness, and temperature, based on the qualitative form of the Landau-level spectrum in an effective four-band model.
\end{abstract}

\maketitle

\section{Introduction}
The study of topological states of matter has undergone a vertiginous growth since the theoretical prediction
\cite{KaneMele2005PRL95-14,BernevigZhang2006,BernevigEA2006} and the experimental observation \cite{KonigEA2007}
of the quantum spin Hall (QSH) effect. Unlike the quantum Hall (QH) effect, which is generated by an external magnetic field, or the quantum anomalous Hall (QAH) effect, which requires time-reversal symmetry (TRS) to be spontaneously broken without applying an external magnetic field, the QSH state is characterized by TRS and is generally driven by the intrinsic spin-orbit (ISO) coupling.\cite{HasanKane2010} Nevertheless, it has been recently shown that in absence of spin-flip terms the QSH effect survives even if the TRS is broken. This state has been dubbed a \emph{weak QSH} state\cite{GoldmanEA2012} (or \emph{TRS broken QSH} state\cite{YangEA2011PRL}), where the weakness refers to the absence of protection by TRS. Indeed, the gap and the topological Chern and spin Chern numbers associated with the topological phase remain robust if the TRS is broken by an exchange term \cite{YangEA2011PRL} or by an additional magnetic field,\cite{GoldmanEA2012,AbaninEA2011,Prodan2009} and there is a quantum phase transition to a topologically distinct or to a trivial phase only when the gap is closed.\cite{YangEA2011PRL,GoldmanEA2012}

An interesting open question is what kind of competition could originate in systems where there is an externally applied magnetic field in addition to  intrinsic magnetic moments, which on their own would lead to the QAH effect.
Recently, the QAH effect has been studied thoroughly for several theoretical models of two-dimensional topological insulators, including HgTe quantum wells\cite{LiuEA2008PRL101} and thin films of Bi${}_2$Se${}_3$\cite{YuEA2010,ChoMoore2011,JinEA2011} doped with transition metal elements such as Mn, Fe, or Cr. In addition, graphene has been proposed as a candidate for the observation of this effect.\cite{QiaoEA2010,*TseEA2011,*ChenEA2011,*TaillefumierEA2011,YangEA2011PRL}
The influence of a magnetic field on Mn-doped HgTe quantum wells has been partially investigated in Ref.~\onlinecite{LiuEA2008PRL101}, with the aim of polarizing the Mn magnetic moment to be eventually able to generate the QAH effect upon shutting down the magnetic field. However, here we concentrate on aspects not considered so far. Since for a certain range of parameters the Zeeman coupling can have a similar effect on the Landau-level (LL) spectrum as the ISO interaction\cite{GoldmanEA2012,AbaninEA2011,Goerbig2011RMP,BeugelingEA2012inprogress} and can also lead to the (TRS broken) QSH effect, it is interesting to explore the interplay between the usual Zeeman term, which is linear in the applied magnetic field, and the non-linear effect arising from the exchange coupling with the magnetic moment of the Mn atoms. In this paper, we show that within a model with a spin-conserving Hamiltonian, the TRS broken QSH phase occurs, and that a \emph{reentrant} behavior is present for a certain range of parameters.

A reentrant integer QH effect has been experimentally observed a few years ago in GaAs quantum wells for filling factors between $\nu = 3$ and $\nu = 4$, in the first LL.\cite{EisensteinEA2002} The phenomenon was later understood to occur due to a sequence of first-order quantum phase transitions between electron-solid (Wigner crystal or bubble phases) and electron-liquid phases,\cite{GoerbigEA2003,*GoerbigEA2004PRB69-11} and it was grounded on the strong electron-electron interactions that dominate the physics at non-integer filling factors. Due to the self-similarity of the Hall conductance curve,\cite{Smet2003,*GoerbigEA2004EPL} which displays a fractal behavior, a similar phenomenon was predicted to occur also for a second-generation of composite fermions.\cite{GoerbigEA2004PRB69-15} In that case, a series of reentrant plateaus would turn out to be quantized at the nearby fractional Laughlin values.\cite{GoerbigEA2004PRL}

\par A second possibility is to observe reentrant integer QH effects solely due to the LL structure of the system. For instance, in Si/SiGe heterostructures, reentrant behavior can be driven at the single-particle level by varying the in-plane magnetic field while keeping the perpendicular field fixed, as to modify the ratio between the cyclotron energy and the Zeeman splitting.\cite{ZeitlerEA2001} In these systems, the crossings of the LLs for spin up and spin down are responsible for the reentrant behavior of the quantized Hall conductivity. For HgTe\cite{KonigEA2008} and InAs/GaSb\cite{NicholasEA2000,*KnezDu2011} quantum wells, the reentrance of the Hall conductivity has been used as a practical method to prove the existence of such a LL crossing and consequently the inverted order of the bands.

In this paper, we show that by applying a magnetic field perpendicular to a Mn-doped HgTe quantum well, the charge and spin Hall conductivities may reenter concomitantly, i.e., there can be a reentrance of the same topological phase, characterized by both its charge and spin topological invariants. This effect is caused by the nonmonotonic behavior of the LL energies due to the nonlinear dependence of the Zeeman term on the externally applied magnetic field. A rich panoply of LL crossings, combined with the nonmonotonicity of the LL energies, provides us with regimes of parameters where this reentrant behavior could be experimentally accessed.
Hg$_{1-Y}$Mn$_{Y}$Te quantum wells are ideal candidates for the observation of these effects, because they have a strong ISO coupling and a large Zeeman $g$-factor. We use the effective four-band model of Ref.~\onlinecite{BernevigEA2006} to compute the LL spectrum\cite{KonigEA2008} together with the relevant Chern numbers in order to identify the QSH state and other QH-like states. Band structure calculations are performed for different values of the quantum well thickness and of the Mn doping fraction to set realistic parameters for this model. We then study the LL spectra including the charge and spin Hall conductivities and determine the conditions for the observation of reentrant topological phases.

\par The outline of this paper is as follows. In Sect.~\ref{sect_model}, we define the effective model that we use to derive our results. In Sect.~\ref{sect_results}, we compute the LL spectrum, explain the mechanisms that lead to the reentrant effects, and explore the parameter regimes for which they can be observed. We conclude by discussing in Sect.~\ref{sect_discussion} the possibilities to resolve the reentrant effects in experiments.

\section{The model}
\label{sect_model}

HgTe and related materials have a zincblende lattice structure, so that the physics of the low-energy electronic bands is well described by the eight-band Kane model.\cite{Kane1957,NovikEA2005} By using perturbation theory (see the Appendix for details), higher-energy bands are projected out in order to reduce this model to an effective four-band model.\cite{BernevigEA2006,NovikEA2005} In this reduced model, the bands under consideration are referred to as $\ket{E_1+},\ket{H_1+},\ket{E_1-},\ket{H_1-}$, in this order. Here, $E$ and $H$ refer to electron- and hole-like bands, respectively, and $+$ and $-$ distinguish the two members of each of the two Kramers pairs $\ket{E_1\pm}$ and $\ket{H_1\pm}$, hereafter referred to as spin components. The symmetry properties under the parity and time-reversal transformations dictate the quadratic-order Hamiltonian $H=H_0+H_\mathrm{Z}+H_\mathrm{ex}$, with\cite{BernevigEA2006,KonigEA2007,KonigEA2008}
\begin{equation}\label{eqn_hamiltonian_full}
  H_0=\begin{pmatrix}h(\vk)&0\\0&h^*(-\vk)\end{pmatrix},
\end{equation}
where
\begin{equation}\label{eqn_hamiltonian_sub}
\begin{aligned}
  h(\vk)&=\epsilon(\vk)\idmat_{2}+d_\alpha(\vk)\sigma^\alpha,\ \ %
  \epsilon(\vk)=C-D(k_x^2+k_y^2),\\
  d_\alpha(\vk)&=(Ak_x,-Ak_y,M(\vk)),\ \ %
  M(\vk)=M-B(k_x^2+k_y^2).
\end{aligned}
\end{equation}
Here, $\sigma^\alpha$ denotes the Pauli matrices and $M$, $A$, $B$, $C$, and $D$ are parameters that depend on the material composition and on the thickness of the quantum well.  In particular, the variations of these parameters induce the topological phase transition from a regime where the electronic bands are ordered normally to a regime where the order is inverted and where the QSH effect is present.\cite{BernevigEA2006}

\par The system is subjected to a perpendicular magnetic field $\magnB\unitvec{z}$, which we will express in terms of the dimensionless variable $\phi$, which denotes the magnetic flux per unit cell measured in units of the flux quantum $h/e$. With these definitions, $\phi$ relates to the magnetic field $\magnB$ and to the magnetic length $\lB$ as $2\pi\phi=e\magnB a^2/\hbar=a^2\lB^{-2}$. For HgTe, with lattice constant $a=0.646\,\mathrm{nm}$, the flux value $\phi=10^{-3}$ corresponds to a magnetic field of $\magnB=9.91\,\mathrm{T}$. In the remainder of this text, we set $C=0$ for convenience, and we set $a\equiv1$ as the unit of length, so that $M$, $A$, $B$, and $D$ all have the dimension of energy.

\par The materials under consideration show a large Zeeman effect, with Land\'e $g$ factors of the order of $20$.\cite{KonigEA2008} We therefore consider the Zeeman term in the Hamiltonian, with different $g$ factors for electrons and holes,
\begin{equation}\label{eqn_hamiltonian_zeeman}
  H_\mathrm{Z}=
  \diag(\tildegE,\tildegH,-\tildegE,-\tildegH)(2\pi\phi)
\end{equation}
where $\tilde{g}_{\mathrm{E}(\mathrm{H})}=g_{\mathrm{E}(\mathrm{H})}\muB\hbar/ea^2\approx g_{\mathrm{E}(\mathrm{H})} \times 91.30\,\mathrm{meV}$ is a rescaled Zeeman parameter, proportional to the Bohr magneton $\muB$ and to the $g$ factor $g_{\mathrm{E}(\mathrm{H})}$ for electrons (holes).\cite{KonigEA2008}

\par In quantum wells of HgTe doped with Mn (with molar fraction $Y$, i.e., we consider Hg$_{1-Y}$Mn$_Y$Te), the presence of Mn has a significant effect on the magnetic properties of the material. It has been found that for low Mn concentrations ($Y\lesssim0.07$), the material behaves paramagnetically, so that its response to the magnetic field is nonlinear: In addition to the Zeeman effect (linear in the magnetic field strength), there is also a nonlinear contribution from the exchange interaction between Mn ions and band states. The exchange interaction term is given by\cite{Furdyna1988,GuiEA2004}
\begin{equation}\label{eqn_hamiltonian_exchange}
  H_\mathrm{ex}=
  \diag(\chiE,\chiH,-\chiE,-\chiH)B_{5/2}(\lambdaex 2\pi\phi)
\end{equation}
where $\chiE$ and $\chiH$ are the exchange energies for the electron and hole bands, respectively,
\begin{equation}\label{eqn_brillouinfunction}
  B_{5/2}(x)=\tfrac{6}{5}\coth(\tfrac{6}{5}x)-\tfrac{1}{5}\coth(\tfrac{1}{5}x)
\end{equation}
is the Brillouin function,\cite{GuiEA2001}
\begin{equation}\label{eqn_lambdaex}
  \lambdaex=\frac{5}{2}\frac{g_\mathrm{Mn}\muB\hbar/ea^2}{\kB(T+T_0)}\approx \frac{5297\,\mathrm{K}}{T+T_0}%
\end{equation}
is an exchange parameter, with $g_\mathrm{Mn}=2$, and $T+T_0$ is an effective temperature, where $T_0\approx 2.6\,\mathrm{K}$.\cite{GuiEA2004} Since the electron wave function $\ket{E_1\pm}$ is a linear combination of the wave functions of the $\Gamma^6$ and $\Gamma^8$ bands, the exchange energy $\chiE$ is a linear combination of the exchange energies $\Delta_s$ and $\Delta_p$ associated with these bands, respectively. The only contribution to the hole wave function $\ket{H_1\pm}$ comes from the $\Gamma^8$ bands, so that $\chiH$ is proportional to $\Delta_p$.\cite{Furdyna1988,LiuEA2008PRL101} For a more detailed explanation, we refer the reader to the Appendix.

\par The energy splitting due to the exchange interactions can be considered as an effective Zeeman splitting, by virtue of the similarity between Hamiltonians~\eqref{eqn_hamiltonian_zeeman} and \eqref{eqn_hamiltonian_exchange}. Here, one writes the Zeeman energy as $\tilde{g}^\mathrm{eff}_{\mathrm{E}(\mathrm{H})}2\pi\phi$, where
\begin{equation}\label{eqn_exchange_geff}
  \tilde{g}^\mathrm{eff}_{\mathrm{E}(\mathrm{H})}(\phi)
   =\tilde{g}_{\mathrm{E}(\mathrm{H})}+\frac{\chi_{\mathrm{E}(\mathrm{H})}}{2\pi\phi}B_{5/2}(\lambdaex 2\pi\phi)
\end{equation}
is the effective, field-dependent $g$ factor for the electron (hole) band.
In the low-field limit ($2\pi\phi\lambdaex\ll 1$), the effective $g$ factor is approximately constant,
$\tilde{g}^{\mathrm{eff}}_{\mathrm{E}(\mathrm{H})}(\phi\to0) = \tilde{g}_{\mathrm{E}(\mathrm{H})}+(7/15)\chi_{\mathrm{E}(\mathrm{H})}\lambdaex$, derived by using a linear approximation to $B_{5/2}(x)$. In the high-field limit $2\pi\phi\lambdaex\gg 1$, the exchange interaction energy is almost constant ($\approx\chi_{\mathrm{E}(\mathrm{H})}$) as a function of the field, because $B_{5/2}(x)\to 1$ for $x\to\infty$,  and as a consequence, it depends also very weakly on the temperature.

\section{Results}
\label{sect_results}
In order to derive the LL spectrum, we model the effect of the magnetic field $\magnB \unitvec{z}$ by the Peierls substitution: In the Hamiltonian, the momentum $\hbar\vk$ is replaced by $\hbar\vk-e\vec{A}$, where $\vec{A}$ is the gauge potential, such that $\magnB\unitvec{z}=\nabla\times\vec{A}$. The freedom of the gauge choice allows us to choose the symmetric gauge, $\vec{A}=(\magnB/2)(-y,x,0)$. Subsequently, we replace $k_+=k_x+\ii k_y$ and $k_-=k_x-\ii k_y$ by the ladder operators $a^\dagger$ and $a$, respectively. These operators raise and lower the LL index by $1$, and their prefactors are chosen such that $[a,a^\dagger]=1$.\cite{KonigEA2008}
In the model presented here, we neglect the coupling between the two spin bands which would arise in the presence of bulk-inversion asymmetry and Rashba spin-orbit coupling. By virtue of this decoupling, the two spin bands can be treated separately. Thus, the eigenvalues and eigenvectors of the Hamiltonian are given by the solutions to the equation $h_\sigma(a,a^\dagger)(\ket{n+1},c\ket{n})=E^{(i)}_{\sigma,n}(\phi)(\ket{n+1},c\ket{n})$, with the appropriate values for $c$. Here, the eigenvalues $E^{(i)}_{\sigma,n}(\phi)$ give the energies of the LLs, where  $n=0,1,2,\ldots$ is the LL index, $\sigma=+,-$ refers to the spin components, and $i=1,2$ distinguishes between the two solutions that exist for each spin component.\cite{KonigEA2008,ButtnerEA2011} For the Hamiltonian that includes the (effective) Zeeman effect, the resulting LL energies are given by\cite{KonigEA2008,ButtnerEA2011}
\begin{widetext}
\begin{align}
  E^{(1,2)}_{\sigma,n}&=[-2n D-\sigma B+\tfrac{1}{2}\sigma\tilde{g}^\mathrm{eff}_+](2\pi\phi)
                        \pm\sqrt{[M+(-\sigma D-2nB+\tfrac{1}{2}\sigma\tilde{g}^\mathrm{eff}_-)(2\pi\phi)]^2+2nA^2(2\pi\phi)}
                        \quad(n\geq1),\nonumber\\
  E_{+,0}&= M -(D+B-\tilde{g}^\mathrm{eff}_\mathrm{E})(2\pi\phi),\qquad\qquad
  E_{-,0} =-M -(D-B+\tilde{g}^\mathrm{eff}_\mathrm{H})(2\pi\phi),\label{eqn_ll_ladder_zeeman}
\end{align}
\end{widetext}
where $\tilde{g}^\mathrm{eff}_\pm\equiv\tilde{g}^{\mathrm{eff}}_{\mathrm{E}}\pm\tilde{g}^{\mathrm{eff}}_{\mathrm{H}}$ are the sum and the difference of the effective (field-dependent) $g$ factors given by \eqn\eqref{eqn_exchange_geff}, including the effect of the Mn doping.

The LL spectra presented here are computed using \eqn\eqref{eqn_ll_ladder_zeeman}, where the relevant parameters have been derived numerically from band structure calculations based on the eight-band Kane model,\cite{Kane1957,NovikEA2005} as explained in more detail in the Appendix. These parameters have been computed for several values of the quantum well thickness $d$ and Mn fraction $Y$. In particular, the dependence of $M$ on $Y$ has dramatic consequences: Increasing $Y$ leads to an increase of $M$, such that it drives the system from the inverted regime ($M<0$) to the topologically trivial regime ($M>0$).\cite{NovikEA2005}

\par The charge (spin) Hall conductivity in a specific bulk gap is defined as the sum of the charge (spin) Chern numbers over all occupied LLs below it. Here, by virtue of the decoupling of the two spin components in the Hamiltonian, the charge and spin Chern numbers of each LL are equal to the sum and difference of the Chern numbers $C_{\pm,n}$ associated with each of the two components. These Chern numbers are well-defined due to the spin-conserving nature of the Hamiltonian. Thus, the charge and spin Hall conductivity expressed in units of their respective quanta, $e^2/h$ and $e/4\pi$, are computed as
\begin{equation}
  \sigmaH=\sum_n(C_{+,n}+C_{-,n}),\quad
  \sigmaspH=\sum_n(C_{+,n}-C_{-,n}),
\end{equation}
where the summation is over the occupied LLs. These values are robust, even in the absence of TRS.\cite{Prodan2009} In this model, each LL contributes a Chern number of $1$, so that the analysis is simplified to merely counting the LLs. The presented values have been verified by analysis of the edge states in a ribbon geometry; see e.g.\ Refs.~\onlinecite{GoldmanEA2012,BeugelingEA2012inprogress} for the details of this alternative approach.

\par The absence of coupling between the two spin states has an important consequence for the QSH phase. Since the QSH state may be viewed as a combination of two independent QH effects for spin up and spin down, it persists even in the absence of time-reversal symmetry.\cite{TkachovHankiewicz2010,YangEA2011PRL}
Additional symmetry-breaking terms, for instance due to bulk-inversion asymmetry and Rashba spin-orbit coupling, would cause an opening of a small gap between the edge states, which allows for some backscattering in the presence of impurities.\cite{YangEA2011PRL,GoldmanEA2012}

\begin{figure*}
  \includegraphics[scale=1]{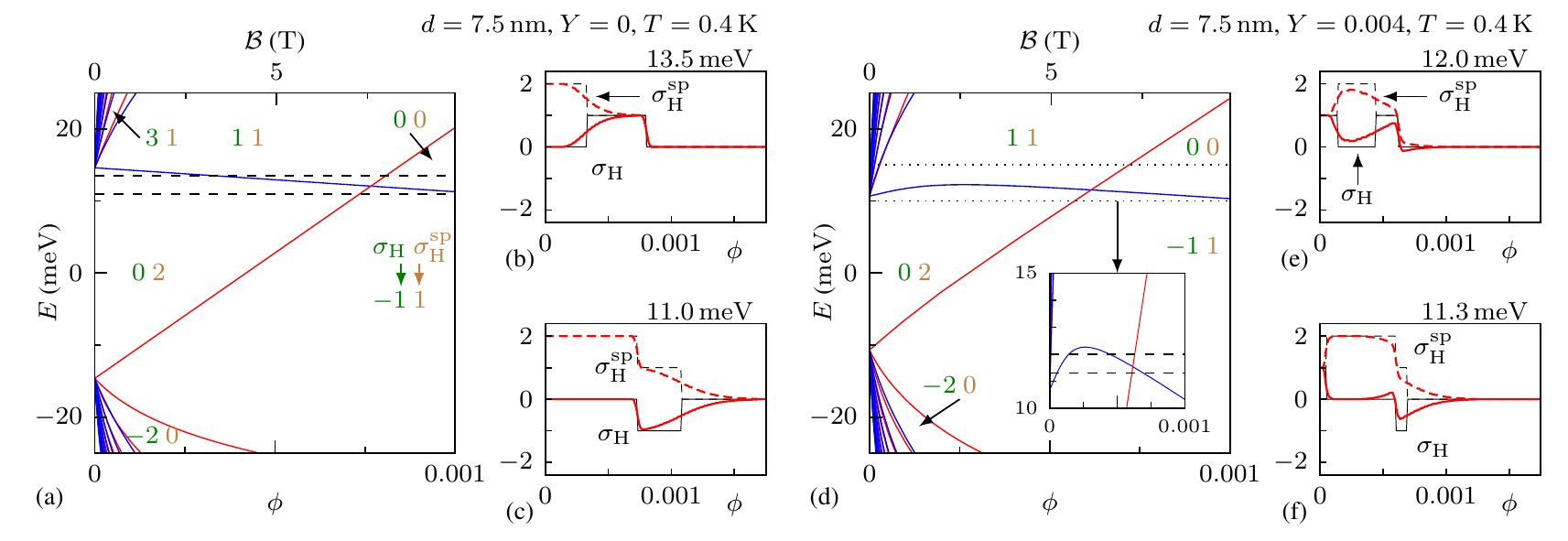}%
  \caption{\label{fig_ll}\coloronline{} (a) LL spectrum for the quantum well with $d=7.5\,\mathrm{nm}$ without Mn doping. Red and blue curves indicate the $+$ and $-$ spin components, respectively. For clarity, only the LLs with $n\leq10$ are shown. The numbers inside the gaps indicate the charge and spin Hall conductivity, $\sigmaH$ and $\sigmaspH$. On the horizontal axes, the flux value $\phi$ and the equivalent magnetic field strength $\magnB$ are given. The dashed lines indicate Fermi energies where reentrant QH effect is observed. (b,c) Charge Hall (solid curves) and spin Hall (dashed curves) conductivity as a function of the flux (in terms of their respective flux quanta) for the indicated Fermi energies. The red (thick) curves indicate the effect of broadening due to disorder ($\Gamma_0=0.3\,\mathrm{meV}$) and temperature ($T=0.4\,\mathrm{K}$), compared to the results without broadening (black, thin curves). (d,e,f) Equivalent plots for $d=7.5\,\mathrm{nm}$ and $Y=0.004$. The inset in (d) is a magnification of the energy region near the LL crossing.}
\end{figure*}

\par In Fig.~\ref{fig_ll}(a), we have displayed the LL spectrum for an undoped ($Y=0$) quantum well with $d=7.5\,\mathrm{nm}$. This system is in the inverted regime, so that the spectrum displays a (weak) QSH gap, (with $(\sigmaH,\sigmaspH)=(0,2)$), for magnetic fields up to $\magnB=7.6\,\mathrm{T}$, where the LLs cross, at $\phi_\mathrm{cross}=M/[2\pi(B-\tilde{g}^\mathrm{eff}_+/2)]$.
In addition to the (weak) QSH gap, we observe several spin-filtered (e.g., $(\sigmaH,\sigmaspH)=(\pm1,1)$), spin-imbalanced (e.g., $(\sigmaH,\sigmaspH)=(3,1)$) and ordinary (e.g., $(\sigmaH,\sigmaspH)=(2,0)$) QH gaps, and a trivial gap ($(\sigmaH,\sigmaspH)=(0,0)$). The (weak) QSH gap is the only gap which exhibits a helical edge state structure; all other nontrivial gaps are chiral. Within this formalism, no other inverted gaps form besides the one at $\phi=0$, because the involved LLs do not cross anywhere else than at $\phi=\phi_\mathrm{cross}$. In contrast, a tight-binding description of a honeycomb lattice in a perpendicular magnetic field does allow for other gaps with helical edge structures at higher flux and Fermi energy values.\cite{GoldmanEA2012,BeugelingEA2012inprogress}

\par The LL spectrum of Fig.~\ref{fig_ll}(a) shows two mechanisms that lead to reentrant behavior of the Hall conductivity and spin Hall conductivity. The first mechanism is illustrated for a Fermi energy of $11.0\,\mathrm{meV}$ (lower dashed line), which lies just below the energy value at which the two lowest Landau levels (LLLs), i.e., the LLs with energies $E_{+,0}$ and $E_{-,0}$, cross. Holding the Fermi energy fixed and increasing the magnetic field, we successively traverse the weak QSH gap with $(\sigmaH,\sigmaspH)=(0,2)$, the spin-filtered QH gap with $(\sigmaH,\sigmaspH)=(-1,1)$, and the trivial gap, where $(\sigmaH,\sigmaspH)=(0,0)$. Thus, the charge Hall conductivity is $0$ for low and high magnetic fields, and $-1$ for intermediate values, which characterizes a reentrance of a charge-insulating state (see Fig.~\ref{fig_ll}(c)). At a Fermi energy slightly above the crossing (e.g., $E=13.5\,\mathrm{meV}$, see Fig.~\ref{fig_ll}(b)), a similar sequence is observed, but with a different intermediate state ($\sigmaH=+1$). In both cases, the spin Hall conductivity takes the values $2$, $1$, and $0$, and does therefore not show reentrant behavior. Clearly the reentrance of the Hall conductivity is caused by the structure of the spectrum around the crossing of the LLLs. To observe the reentrance of the charge Hall conductivity, it is essential that the derivatives $\mathrm{d}E/\mathrm{d}\phi$ of the two LLLs at the crossing differ in sign, which can happen only in the inverted regime. Thus, experimental observation of this type of reentrance provides a proof that the HgTe quantum well can indeed be described as an inverted Dirac system.\cite{KonigEA2007} One may verify that if the signs of the derivatives would be equal, then the charge Hall conductivity does not reenter. Instead, we would observe reentrant spin Hall conductivity. We note that crossings of the latter type are ubiquitous for higher LLs ($n>0$), but they are difficult to observe due to the vicinity of other LLs. However, later we will show that, under some circumstances, the crossings of the LLLs may also be of this type.

\par In Fig.~\ref{fig_ll}(d), we show the effect of doping ($Y=0.004$) on the LL spectrum. Two effects are visible. First, the size of the (weak) QSH gap has decreased, consistent with the increase of $M$. Secondly, the energies $E_{\pm,0}$ of the two LLLs are no longer linear in the magnetic field. In fact, one of these LLLs shows a nonmonotonic dependence on the field. As can be observed in Fig.~\ref{fig_ll}(d), this nonmonotonic LLL attains its maximum for a flux value less than $\phi_\mathrm{cross}$. Thus, if the Fermi energy is located between the energy of the crossing and that of the maximum (e.g., if $E=12.0\,\mathrm{meV}$, see the inset of Fig.~\ref{fig_ll}(d) and Fig.~\ref{fig_ll}(e)), the spin-filtered QH gap reenters, and the intermediate state is the (weak) QSH gap. Thus, the system goes from a chiral, to a helical, and back to the (same) chiral phase again. This simultaneous reentrance of the charge and spin Hall conductivity should be contrasted with the reentrant behavior around the LLL crossing, where only one of them reenters, but not both. We remark that such a sequence is possible only if the intermediate phase is the (weak) QSH phase, and consequently only if the LL involved is one of the LLLs, since the higher LLs are all monotonic. Therefore, this behavior cannot be observed in the undoped system, where the LLL energies are linear.

\par As can be observed in the inset of Fig.~\ref{fig_ll}(d), the maximum of this LLL has an energy close to that of the LL crossing. The sequence of charge and spin Hall conductivities is therefore affected by both mechanisms. We shall call this phenomenon \emph{compound} reentrant behavior. Above the energy of the crossing, the aforementioned sequence (spin-filtered QH, weak QSH, spin-filtered QH) is followed by the trivial gap, so that we get an additional reentrance of the zero charge Hall conductivity. Just below the crossing energy (e.g., $E=11.3\,\mathrm{meV}$, see Fig.~\ref{fig_ll}(f)), the sequence of gaps is spin-filtered QH $(1,1)$, weak QSH $(0,2)$, spin-filtered QH $(-1,1)$, and trivial $(0,0)$. In this situation, the two spin-filtered QH phases are \emph{different} gaps, unlike the sequence above the crossing. These examples show that the rich compound reentrant behavior will appear if the crossing and the maximum of the LLLs are close to each other.

\begin{figure*}
  \includegraphics[scale=1]{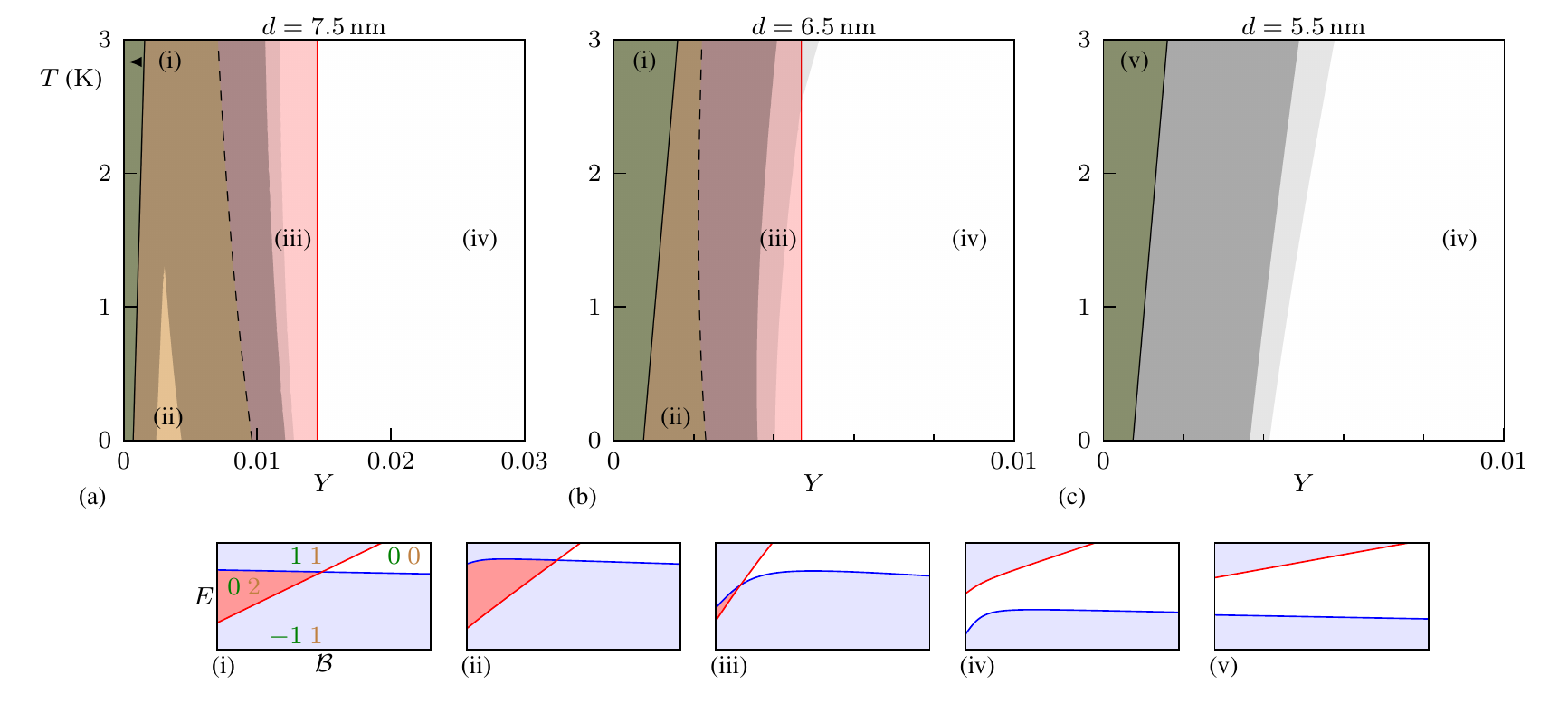}
  \caption{\label{fig_diag_reentrant_types}\coloronline{} Diagram showing the possible types of reentrant behavior, as a function of the Mn fraction $Y$ and temperature $T$. In the top row, we show diagrams for (a) $d=7.5\,\mathrm{nm}$, (b) $d=6.5\,\mathrm{nm}$, and (c) $d=5.5\,\mathrm{nm}$. The roman numbers correspond to different regimes in the qualitative structure of the spectrum constituted by the LLLs, displayed in the bottom row. The characteristics of the regimes (i)--(v) are explained in the text. In the top row, the different shades indicate the energy range in which the compound reentrant behavior is present, compared to the LL broadening: Brighter colors indicate a larger range, making the effect easier to observe. The dashed line indicates the parameter values where the crossing and the maximum of the nonmonotonic LLL coincide, separating (ii) and (iii). The shading in the bottom row indicates the weak QSH (red/dark gray), spin-filtered QH (blue/light gray), and trivial (white) gaps.}
\end{figure*}

\par In order to be able to observe the reentrant effects in experiments, we study the qualitative structure of the LL spectrum as a function of the well width $d$, the doping fraction $Y$, and the temperature $T$. More specifically, for a fixed choice of parameters, we analyze whether one of the LLL is nonmonotonic, and whether the LLLs cross. Furthermore, if the nonmonotonicity and crossing appear at the same time, we determine the position of the maximum/minimum and the crossing relative to each other. For simplicity, we restrict ourselves to the structure of the two LLLs.

\par The bottom row of Fig.~\ref{fig_diag_reentrant_types} displays five qualitatively different LLL spectra, which distinguish the regimes as given by Fig.~\ref{fig_diag_reentrant_types}(a)--(c). These regimes are characterized as follows. For regimes (i)--(iii), the band gap has inverted order (i.e., $M<0$), and therefore shows the QSH phase at zero magnetic field. In regime (i), the LLLs are monotonic, so that the only mechanism that leads to reentrant effects is the crossing. In regimes (ii) and (iii) one LLL is nonmonotonic, so that we have compound reentrant behavior. These two regimes are distinguished by the flux value of the maximum, which is smaller (ii) or greater (iii) than the flux value of the crossing. In the case (iii), both LLLs are increasing at the crossing, so that we observe reentrant spin Hall conductivity, as argued before. In regimes (iv) and (v), the band gap is normally ordered (i.e., $M>0$), so that we find a trivial phase at zero magnetic field. In this situation, the LLLs do not cross, and the only mechanism that can lead to reentrant behavior is the presence of a nonmonotonic LLL, as is the case (iv). For regime (v), both LLL are monotonic and do not cross, thus preventing any type of reentrant behavior.

\section{Discussion}
\label{sect_discussion}
Let us finally comment on the ability to resolve these reentrant effects, based on the range of the Fermi energies for which they are present. In order to estimate the observability, we compare this energy range to the broadening of the LLs, that will cause the change of conductivity across a LL to be smooth rather than step-like. Here, we consider a gaussian broadening with width $\Gamma$,\cite{NovikEA2005} that incorporates both LL broadening due to disorder and the smooth variation of the fermionic filling function at finite temperatures. The broadening due to disorder has a field-dependent width $\Gamma_\mathrm{dis}=\Gamma_0\sqrt{\magnB/\magnB_0}$, where $\Gamma_0\sim0.1$--$2\,\mathrm{meV}$ and $\magnB_0\equiv1\,\mathrm{T}$.\cite{NovikEA2005} The thermal broadening is approximated by a gaussian with width $\Gamma_\mathrm{th}=\sqrt{2/3}\pi\kB T$. In Fig.~\ref{fig_ll}(b,c,e,f), we illustrate the smooth transitions of the conductivities due to the combined effects of both types of broadening.

\par We consider the compound reentrant effects displayed in Fig.~\ref{fig_diag_reentrant_types}(ii) and (iii) to be well resolvable if the difference between the energies at the crossing and at the maximum exceeds twice the broadening width $2\Gamma$. Indeed, in that case, the difference between the actual conductivity values and the quantized ones in absence of broadening effects is $\lesssim0.08$. However, the value of $2\Gamma$ is not a hard limit: Variations of the quantized Hall conductivity, even if they are far from the quantized values, may already be considered as a signature for a reentrant effect, for example as demonstrated in Fig.~\ref{fig_ll}(e).
In the diagrams of Fig.~\ref{fig_diag_reentrant_types}(a)--(c), the different shadings indicate this distance compared to $\Gamma$. In the brightest regions, the distance between the LLs is larger than $2\Gamma$, sufficient for the reentrant effect to be observed. We find that for $\Gamma_0=0.3\,\mathrm{meV}$, situation (ii) is difficult to observe due to the small energy difference between the LLs, whereas the observation of (iii) is easier close to the critical doping, above which the system goes to the trivial regime (iv). For the observation of the compound reentrant effects,  thicker wells are favorable because the energy range where the effects appear is larger. The simple reentrant effect due to nonmonotonicity, as in situation (iv), is generally present in a large energy regime and therefore its observation is less affected by the LL broadening.

\par Transport experiments with HgTe quantum wells have so far concentrated on the charge Hall conductivity of the system. For example, the reentrance of the charge Hall conductivity has been utilized to identify the possible regime of the QSH phase.\cite{KonigEA2008} Observation of the simultaneous reentrant behavior of the charge and spin Hall conductivity would also require the availability of a spin-sensitive detector, e.g., a contact consisting of a tunneling barrier and a ferromagnet.\cite{LouEA2007} However, this technique has the drawback that it only works at low fields, within the hysteresis range of the ferromagnet. Another detection mechanism could be a local magneto-optical Kerr effect (MOKE) experiment,\cite{KatoEA2004} although this measurement would be difficult due to the small bandgap of the semiconductor. The inverse spin Hall effect may provide a way to measure the spin polarization of the edge states.\cite{BruneEA2010,*BruneEA2011preprint-1107} Nevertheless, the measurement of the charge Hall conductivity at multiple Fermi energies together with knowledge of the structure of the spectrum may provide indirect evidence for the existence of these reentrant effects.

\par In conclusion, we have demonstrated that the nonmonotonic behavior of the LLs in the presence of Mn doping leads to reentrant topological phases, and that the vicinity of LL crossings leads to rich compound reentrant behavior. Five different qualitative forms of the structure of the LLLs were shown to occur in the parameter space characterized by Mn doping, well thickness, and temperature. Furthermore, we have investigated the effects of LL broadening to estimate the ability to resolve the reentrant effects in experiments.

\acknowledgments
We thank V. Juri\v{c}i\'{c} and E. M. Hankiewicz for useful discussions.  This work was supported by the Netherlands Organisation for Scientific Research (NWO) (W. B. and C. M. S.), the German Research Foundation DFG [SPP 1285 Halbleiter Spintronik, DFG-JST joint research program (L. W. M.), and Grant No. AS327/2 (E. G. N.), the Alexander von Humboldt Foundation (C. X. L.), and the EU ERC-AG program (L. W. M.).

\appendix*
\section{Numerical methods and derivation of the four-band effective model by perturbation theory}
\label{ch_appendix}%
In this appendix, we illustrate the used numerical method and relate it to the perturbation theory which allows us to determine the parameters of a four-band effective model. In the Kane model, the band structure of the material consists of eight bands.\cite{Kane1957} However, the two bands $|\Gamma^7,\pm1/2\rangle$ are separated by approximately $1\,\mathrm{eV}$ from the other six bands and will be neglected here. The resulting six-band modified Kane Hamiltonian is written in the basis $|\Gamma^6,1/2\rangle$, $|\Gamma^6,-1/2\rangle$,
$|\Gamma^8,3/2\rangle$, $|\Gamma^8,1/2\rangle$,
$|\Gamma^8,-1/2\rangle$ and $|\Gamma^8,-3/2\rangle$, which we denote
as $|1\rangle$, $|2\rangle$, $|3\rangle$,
$|4\rangle$, $|5\rangle$ and $|6\rangle$ for short in the following. The Hamiltonian can then be written as
\begin{equation}\label{eqn_appHam1}
    H=H_0+H_\mathrm{Z}+H_\mathrm{ex},
\end{equation}
where $H_0$ is the six-band Kane Hamiltonian,\cite{Kane1957,RotheEA2010} $H_\mathrm{Z}$ is the linear Zeeman term and $H_\mathrm{ex}$ is due to the exchange interaction between the Mn ions and the band states in a magnetic field $\magnB$ in the $z$ direction. The Zeeman term reads
\begin{align}
    H_\mathrm{Z}^\mathrm{c}&=\frac{g_0}{2}\muB\magnB\sigma_z,\nonumber \\
    H_\mathrm{Z}^\mathrm{v}&=\kappa'\muB\magnB \hat{J}_z,
    \label{eqn_appHamZeeman}
\end{align}
for the (decoupled) conduction ($|1\rangle$ and $|2\rangle$) and valence ($|3\rangle$,
$|4\rangle$, $|5\rangle$, and $|6\rangle$) band parts of the Hamiltonian. Here $\hat{J}_z$ is the angular momentum operator, $g_0$ is the bare Zeeman $g$-factor of HgTe, and $\kappa'$ is a phenomenological parameter.\cite{NovikEA2005}
The exchange term, induced by the $sp$-$d$ coupling between the Mn $d$ level electrons and
conduction or valence band electrons, has a similar form as the Zeeman term, and reads
\begin{align}
    H_\mathrm{ex}^\mathrm{c}=-\Delta_s\sigma_z\nonumber\\
    H_\mathrm{ex}^\mathrm{v}=-\frac{2}{3}\Delta_p\hat{J}_z
    \label{eqn_appHamex}
\end{align}
where $\Delta_s=0.2\,\mathrm{eV}\times Y\avg{\mathcal{S}}$ and $\Delta_p=-0.3\,\mathrm{eV}\times Y\avg{\mathcal{S}}$
are the coupling constants between the Mn spin $\vec{\mathcal{S}}$ and
the conduction band ($\Delta_s$) or the valence band ($\Delta_p$),
respectively, and $Y$ is the mole fraction of Mn$^{2+}$ ions. The polarization of the Mn spin
$\vec{\mathcal{S}}$ is assumed to be in the $z$ direction. We
regard the Mn spin as a classical spin and use the mean field value
$\avg{\mathcal{S}}$ instead of $\vec{\mathcal{S}}$, which yields
\begin{equation}\label{eqn_appbrillouin}
  S\equiv\avg{\mathcal{S}}=-S_0B_{5/2}\left(\frac{5g_\mathrm{Mn}\muB\magnB}{2\kB(T+T_0)}\right),
\end{equation}
where $B_{5/2}$ is the Brillioun function as given by \eqn\eqref{eqn_brillouinfunction}, $S_0=5/2$, $g_\mathrm{Mn}=2$, and $T_0\approx 2.6\,\mathrm{K}$ for Mn.\cite{GuiEA2004} The argument of $B_{5/2}$ in this equation is equal to $\lambdaex 2\pi\phi$, cf.~\eqns\eqref{eqn_hamiltonian_exchange} and \eqref{eqn_lambdaex}.

Now, we consider the above model in a periodic superlattice grown in the
$z$-direction with well width $d$ and barrier width $L-d$. In the limit of large $L-d$, it becomes equivalent to a single quantum well. Due
to the periodic boundary condition along the $z$-direction, according to
Bloch's theorem, we can write the wave function as
\begin{equation}
    \Psi_{\xi}=\frac{1}{2\pi}\ee^{\ii(\vec{k}_{\parallel}\cdot\vec{r}_\parallel+k_zz)}
    |U^{\xi}_{\vec{k}}(z)\rangle,
    \label{eqn_PsiWf}
\end{equation}
where $\vec{k}=(\vec{k}_{\parallel},k_z)=(k_x,k_y,k_z)$ and $(\vec{r}_\parallel,z)=(x,y,z)$. The in-plane wave vector $\vec{k}_{\parallel}$ is a good quantum number for the system, and $k_z$ is the superlattice wave number in the $z$ direction, taken to be zero, because the quantum wells are effectively decoupled for large barrier thickness $L-d$. $U^{\xi}_{\vec{k}}(z)$ is a multi-component periodic wave function
$U^{\xi}_{\vec{k}}(z+L)=U^{\xi}_{\vec{k}}(z)$ of the $\xi$-band, which is expanded in terms of a plane-wave basis as
\begin{equation}
    |U^{\xi}_{\vec{k}}(z)\rangle=\sum_{n,\lambda}a^{\xi}_{n,\lambda}|n,\lambda\rangle
    =\sum_{n,\lambda}a^{\xi}_{n,\lambda}\frac{1}{\sqrt{2\pi}}\ee^{\ii (2\pi n/L)z}|\lambda\rangle,
    \label{eqn_Uwf}
\end{equation}
where $|\lambda\rangle$ denotes the component $\lambda=1,\ldots,6$ of the wave function, and the expansion coefficients $a^{\xi}_{n,\lambda}$ are functions of $\vec{k}_{\parallel}$. The eigenequation for these states is given by
$
    \hat{H}\Psi_{\xi}=E_{\xi}\Psi_{\xi},
$
where $E_{\xi}$ depends on $\vec{k}_{\parallel}$. With the expansion \eqref{eqn_Uwf}, we find
\begin{equation}
    \sum_{n',\lambda'}\langle n,\lambda|\hat{H}|n',\lambda'\rangle a^{\xi}_{n',\lambda'}
    =E_{\xi}a^{\xi}_{n,\lambda}.
    \label{eqn_EigenEqn2}
\end{equation}
A truncation method is applied and a finite number of basis vectors ($n=-N,-N+1,\ldots,N-1,N$) is used to solve this eigenvalue problem to obtain the coefficients $a^{\xi}_{n,\lambda}$. Given the fact that we are only interested in the low-energy physics, taking $N=20$ yields a solution that is sufficiently accurate.

\begin{figure}[t]
\includegraphics[width=86mm] {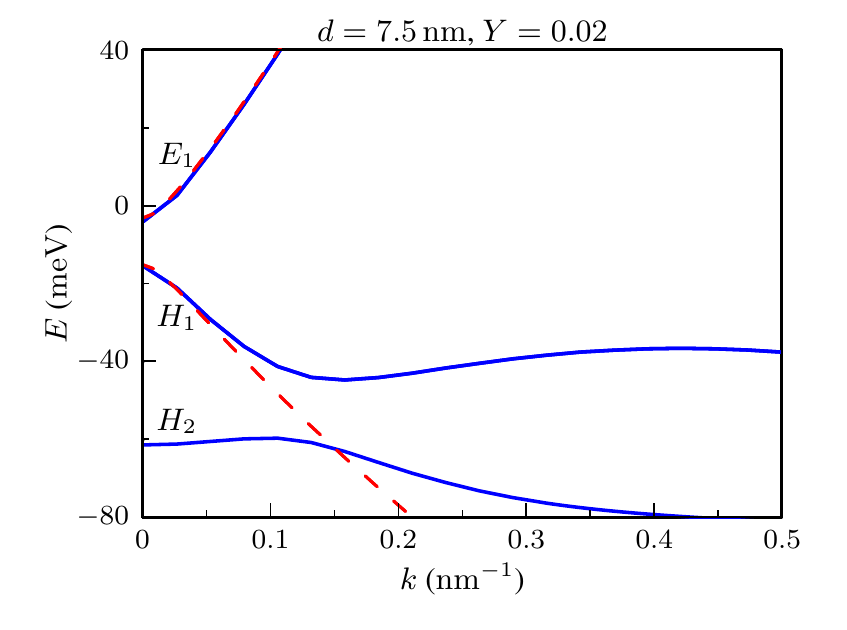}
\caption{\label{fig_DispCom}\coloronline{} Comparison of the energy dispersion calculated from the full
Hamiltonian (blue solid curves) and from the effective model (red dashed curves) for $d=7.5\,\mathrm{nm}$ and $Y=0.02$. The two results exhibit a good overlap in the low-energy regime, which demonstrates the reliability of the effective
model.}
\end{figure}

Next, we relate the perturbation theory to the previous
numerical method. The Hamiltonian \eqref{eqn_appHam1} is divided into
\begin{equation}
    H=H_{\vec{k}_{\parallel}=0}+H_{\vec{k}_{\parallel}}^{(1)},
    \label{eqn_Hdivide}
\end{equation}
where $H_{\vec{k}_{\parallel}=0}$ is treated as the zero-order Hamiltonian and $H_{\vec{k}_{\parallel}}^{(1)}$ as
the perturbation. The wave function at the $\Gamma$ point
($\vec{k}_{\parallel}=0$) can be obtained from the numerical
calculation, which is denoted as
\begin{equation}
    |U^{\xi}_{\vec{k}=0}(z)\rangle=\sum_{\lambda}f_{\xi,\lambda}(z)|\lambda\rangle
    \label{eqn_Uwf2}
\end{equation}
with $f_{\xi,\lambda}(z)=\sum_na^{\xi}_{n,\lambda}|n\rangle$. Only the subbands
$|E_1+\rangle,|H_1+\rangle,|E_1-\rangle,|H_1-\rangle$, which are
denoted as $|A\rangle,|B\rangle,|C\rangle,|D\rangle$ for short, are
concerned in this calculation. Using symmetry arguments, we obtain
\begin{align}
|A\rangle&=f_{A,1}(z)|1\rangle+f_{A,4}(z)|4\rangle,\quad&
|B\rangle&=f_{B,3}(z)|3\rangle,\nonumber\\
|C\rangle&=f_{C,2}(z)|2\rangle+f_{C,5}(z)|5\rangle,\quad&
|D\rangle&=f_{D,6}(z)|6\rangle,
\end{align}
where $f_{A,1}=f_{C,2}$,$f_{A,4}=f_{C,5}$,$f_{B,3}=f_{D,6}$. Under
two-dimensional spatial reflection,
$f_{A,1},f_{C,2},f_{B,3},f_{D,6}$ have even parity and
$f_{A,4},f_{C,5}$ have odd parity. Furthermore, in order to take
into account the contribution of the other subbands in
second-order perturbation theory, additional states $|E_2\pm\rangle$, $|LH\pm\rangle$,
$|HH_2\pm\rangle$, and $|HH_3\pm\rangle$ (the second electron, light hole, and second and third heavy hole bands, respectively) are also solved numerically and can be written in a similar way.\cite{RotheEA2010}

\begin{table*}[t]
\caption{\label{tbl_parameters} Parameters for the four-band effective model [\eqns\eqref{eqn_hamiltonian_full}--\eqref{eqn_hamiltonian_exchange}], obtained by perturbation theory from the full Kane model.}
\begin{ruledtabular}%
\begin{tabular}{dd|ddddddddd}
\multicolumn{1}{c}{\mbox{$d$ ($\mathrm{nm}$)}} &
\multicolumn{1}{c|}{\mbox{$Y$}} &
\multicolumn{1}{c}{\mbox{$C$ ($\mathrm{meV}$)}}&
\multicolumn{1}{c}{\mbox{$M$ ($\mathrm{meV}$)}}&
\multicolumn{1}{c}{\mbox{$A$ ($\mathrm{eV}$)}} &
\multicolumn{1}{c}{\mbox{$B$ ($\mathrm{eV}$)}} &
\multicolumn{1}{c}{\mbox{$D$ ($\mathrm{eV}$)}} &
\multicolumn{1}{c}{\mbox{$g_\mathrm{E}$}}      &
\multicolumn{1}{c}{\mbox{$g_\mathrm{H}$}}      &
\multicolumn{1}{c}{\mbox{$F_1$}} &
\multicolumn{1}{c}{\mbox{$F_4$}} \\
\hline
5.5 & 0.00 & -16.9 &   8.8 &  0.60 & -1.15 & -0.73 &  15.8 &  1.22 &  0.62 &  0.37\\
5.5 & 0.01 &  -5.8 &  20.0 &  0.62 & -1.05 & -0.63 &  14.4 &  1.29 &  0.64 &  0.35\\
5.5 & 0.02 &   5.6 &  31.5 &  0.64 & -0.96 & -0.55 &  13.2 &  1.36 &  0.66 &  0.33\\
5.5 & 0.03 &  17.4 &  43.3 &  0.66 & -0.89 & -0.48 &  12.2 &  1.42 &  0.68 &  0.31\\
6.5 & 0.00 & -24.4 &  -4.9 &  0.58 & -1.45 & -1.04 &  20.0 &  1.22 &  0.58 &  0.41\\
6.5 & 0.01 & -13.9 &   5.7 &  0.60 & -1.30 & -0.88 &  18.0 &  1.28 &  0.61 &  0.38\\
6.5 & 0.02 &  -3.0 &  16.6 &  0.62 & -1.17 & -0.75 &  16.3 &  1.35 &  0.63 &  0.36\\
6.5 & 0.03 &   8.4 &  28.0 &  0.65 & -1.06 & -0.65 &  14.9 &  1.42 &  0.66 &  0.34\\
7.5 & 0.00 & -29.9 & -14.6 &  0.55 & -1.87 & -1.45 &  24.3 &  1.21 &  0.55 &  0.44\\
7.5 & 0.01 & -19.9 &  -4.6 &  0.58 & -1.62 & -1.20 &  21.8 &  1.28 &  0.58 &  0.41\\
7.5 & 0.02 &  -9.5 &   5.8 &  0.61 & -1.42 & -1.00 &  19.5 &  1.34 &  0.61 &  0.39\\
7.5 & 0.03 &   1.4 &  16.8 &  0.63 & -1.26 & -0.85 &  17.6 &  1.41 &  0.64 &  0.36
\end{tabular}%
\end{ruledtabular}%
\end{table*}

\par With the obtained zero-order wave function, we apply the second-order perturbation formalism\cite{RotheEA2010}
\begin{multline}
    H_{m'm}=\langle m'|H|m\rangle+\sum_{s}\frac{1}{2}
    \langle m'|H_{\vec{k}_{\parallel}}^{(1)}|s\rangle\langle
    s|H_{\vec{k}_{\parallel}}^{(1)}|m\rangle\\
    {}\times\left(\frac{1}{E_{m'}-E_s}+\frac{1}{E_m-E_s}\right),
\end{multline}
to obtain the effective model given by \eqns\eqref{eqn_hamiltonian_full}--\eqref{eqn_hamiltonian_exchange}. Here $|m\rangle,|m'\rangle$ are the states chosen
from $|A\rangle$, $|B\rangle$, $|C\rangle$, and $|D\rangle$ while
$|s\rangle$ is one of the intermediate states $|E_2\pm\rangle$, $|LH\pm\rangle$,
$|HH_2\pm\rangle$, and $|HH_3\pm\rangle$. With this approach, we
relate the parameters of the effective model \eqref{eqn_hamiltonian_full}--\eqref{eqn_hamiltonian_exchange} to the parameters of the six-band modified Kane model
\eqref{eqn_appHam1}--\eqref{eqn_appHamex}. We find that for the effective mass
parameters $D$ and $B$ and for the effective $g$ factor $g_\mathrm{E}$,
we need to take into account the second-order perturbation, while
for the other parameters the first-order term is accurate enough
for our purpose. As the derivation is straightforward and the
expressions for the parameters are quite lengthy, we do not write them explicitly
here. As an example, the exchange parameters $\chiE$ and $\chiH$ are given by
\begin{align}
    \chi_\mathrm{E}B_{5/2}(\lambdaex 2\pi\phi)&=-(F_1\Delta_s+F_4\Delta_p/3),\nonumber\\
    \chi_\mathrm{H}B_{5/2}(\lambdaex 2\pi\phi)&=-\Delta_p,
    \label{eqn_app_chiEH}
\end{align}
where $F_1=\langle f_{A,1}|f_{A,1}\rangle$ and $F_4=\langle
f_{A,4}|f_{A,4}\rangle$, and $\lambdaex 2\pi\phi$ is the argument of $B_{5/2}$ in \eqns\eqref{eqn_hamiltonian_exchange} and \eqref{eqn_appbrillouin}.
In Table~\ref{tbl_parameters}, we show the numerical values
of these parameters for several different well thicknesses and
different Mn doping. In Fig.~\ref{fig_DispCom}, the energy dispersion
calculated from the effective model using the parameters in Table~\ref{tbl_parameters} is
shown to fit well with that calculated from the full Kane model at
small $k$. This result justifies the use of the effective model to
discuss the low-energy physics, in particular in the energy range where the reentrant behavior occurs.
In this paper, we have restricted ourselves to wells with a thickness $d<8.1\,\mathrm{nm}$, because above this value, the $H_2$ band lies between the $E_1$ and $H_1$ bands, and in that case the four-band model is no longer accurate, especially in the energy regime of the valence band. Nevertheless, the mechanisms for appearance of the reentrant effects may still be present for thicker wells.


%

\end{document}